\documentclass[prl,showpacs,showkeys,twocolumn]{revtex4}
\usepackage{makeidx}
\usepackage{amsmath}
\usepackage{eurosym}
\usepackage{amssymb}
\usepackage{graphicx}
\usepackage{dcolumn}
\usepackage{bm}
\usepackage{rotating}

\begin{document}

\title{Intervalley-Scattering Induced Electron-Phonon Energy Relaxation in
Many-Valley Semiconductors at Low Temperatures}
\author{M. Prunnila$^1$}
\email{mika.prunnila@vtt.fi}
\author{P. Kivinen$^2$}
\author{A. Savin$^3$}
\author{P. T\"{o}rm\"{a}$^{2}$}
\author{J. Ahopelto$^{1}$}
\affiliation{$^{1}$VTT Information Technology, P.O.Box 1208, FIN-02044 VTT, Espoo,
Finland }
\affiliation{$^{2}$NanoScience Center, Department of Physics, University of Jyv\"{a}skyl%
\"{a}, P.O.Box 35, FIN-40014 Jyv\"{a}skyl\"{a}, Finland}
\affiliation{$^3$Low Temperature Laboratory, Helsinki University of Technology, P.O.Box
2200, FIN-02015 HUT, Finland}
\date{\today}

\begin{abstract}
We report on the effect of elastic intervalley scattering on the energy
transport between electrons and phonons in many-valley semiconductors. We
derive a general expression for the electron-phonon energy flow rate at the
limit where elastic intervalley scattering dominates over diffusion.
Electron heating experiments on heavily doped n-type Si samples with
electron concentration in the range $3.5-16.0\times 10^{25}$ m$^{-3}$ are
performed at sub-1 K temperatures. We find a good agreement between the
theory and the experiment.
\end{abstract}

\keywords{electron-phonon interaction, energy relaxation, disordered
semiconductors }
\pacs{63.20.Kr, 44.90.+c}
\maketitle

%\preprint{APS/123-QED}

%\email{Pasi.Kivinen@phys.jyu.fi}

%\email{mika.prunnila@vtt.fi}

%\textbf{Keywords:} Quantum well, Silicon-on-Insulator, Quantum Hall, Electron Transport

%\textbf{Keywords:} Quantum well, Silicon-on-Insulator, Quantum Hall, Electron Transport

%72. Electronic transport in condensed matter  
%71.70.Di Landau levels (condensed matter)  
%73.21.Fg Quantum wells (electron states/collective excitations) 
%73.43.-f Quantum Hall effects 

%\pagestyle{myheadings} \markright{Submitted to Solid-State Electronics}

%\thispagestyle{myheadings} \markright{Submitted to Solid-State Electronics}

%\thispagestyle{myheadings}
%\markright{Submitted to Appl. Phys. Lett.}
%$^{\text{a)}}$}

%\author{$^{\text{a)}}$Electronic mail: mika.prunnila@vtt.fi}

%\date{\today\footnote{Submitted to Appl. Phys. Lett}}}

%\preprint{APS/123-QED}

% Force line breaks with \\

%\email{mika.prunnila@vtt.fi}

% \email{Second.Author@institution.edu}

% It is always \today, today,
%  but any date may be explicitly specified

% PACS, the Physics and Astronomy
% Classification Scheme.
%\keywords{electron mobility, elastic scattering, silicon-on-insulator, quantum well}}%Use showkeys class option if keyword
%display desired

%\kword{electron mobility, elastic scattering, silicon-on-insulator, quantum well}

Since the low temperature hot electron experiments by Roukes \textit{et al. }%
\cite{roukes:1985}, the energy transport between electrons and phonons has
continued to be a topical subject. Recently, there has been significant
experimental and theoretical interest in the electron-phonon (e-ph) energy
relaxation in metals and semiconductors at low temperatures \cite%
{sergeev:2000,agan:2001,gershenson:2001,prus:2002,kivinen:2003,meikap:2004,sergeev:2005}%
. The understanding of thermal e-ph coupling is important for several low
temperature devices such as microbolometers, calorimeters and on chip
refrigerators \cite{gershenson:2001,anghel:2001}. This coupling plays also
an important role in correct interpretation of low temperature experiments 
\cite{prus:2002} and the e-ph energy relaxation rate gives direct
information about phonon mediated electron dephasing \cite{bergmann:1990}.

Interaction between electrons and phonons is strongly affected by the
disorder of the electron system and, therefore, the problem is commonly
divided into two special cases: pure and impure (or diffusive) limit of e-ph
interaction. The cross-over between these two regions is defined as $ql=1$ ,
where $q$ is the phonon wavevector and $l$ the electron mean free path. If
the whole phonon system is to be considered then the phonon wavevector can
be conveniently replaced by the thermal phonon wave vector $%
q_{T}=k_{B}T/\hbar v$, where $T$ is the temperature of the lattice and $v$
the sound velocity. \ Recent theory for single-valley semiconductors \cite%
{sergeev:2005} predicts that the e-ph energy relaxation is strongly enhanced
when the system enters from the pure limit ($ql>1$) to the diffusive limit ($%
ql<1$). The behavior is the opposite in comparison to metals where it is
well known, since the pioneering work by A.\ B. Pippard \cite{pippard:1955},
that the disorder of the electron system tends to suppress the e-ph energy
relaxation (see also Ref. \cite{sergeev:2000}). In semiconductors, due to
small electron density, the e-ph interaction can be described by deformation
potential coupling constants, which do not depend on the electronic
variables, while in metals the coupling strongly depends on the electron
momentum \cite{khan:1984}. This fundamental difference eventually leads to
disorder enhancement of the relaxation in the diffusive limit in
single-valley semiconductors \cite{sergeev:2005}. 
\begin{figure}[b]
\begin{center}
\includegraphics[width=50mm,height=!]{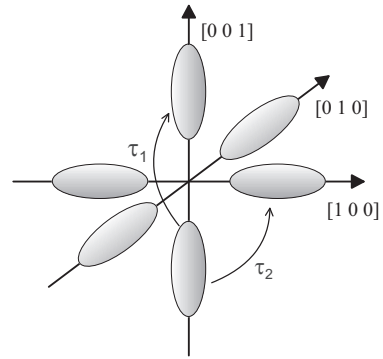}
\end{center}
\caption{Schematic illustration of the constant energy ellipsoids of Si
conduction band valleys. The valleys are located close to X-point in the
first Brillouin zone. Elastic scattering rates $1/\protect\tau _{1}$ and $1/%
\protect\tau _{2}$ couple the different classes of the valleys.}
\label{Si_band}
\end{figure}

In many-valley semiconductors the situation is further altered due to
intervalley scattering, which is the topic of our work. Due to lack of
screening the e-ph energy flow rate is strongly enhanced in many valley
semiconductors in comparison to single valley ones at diffusive low
temperature limit. We approach the e-ph energy transport problem by first
considering the phonon energy attenuation \ rate due to electrons (or
phonon-electron energy relaxation rate). This procedure is attractive,
because it enables straigth forward comparison between our work and previous
literature, which has concentrated mainly on ultrasonic attenuation \cite%
{weinreich:1959,dutoit:1970,sota:1982,sota:1986,sachdev:1987}. We derive
expression for the total e-ph energy flow rate (by using the phonon energy
attenuation rate) and perform low temperature electron heating experiments
to heavily doped n-type silicon samples. We find excellent agreement between
the theoretical and the experimental e-ph temperature responses.

As discussed above the electron-phonon coupling in semiconductors can be
described through deformation potential coupling constants, which do not
depend on the electron variables (in a single valley). The strain induced
conduction band energy shifts $\delta v_{l}$ ($l=1,2,\ldots ,L$, $\ \ L$ is
the number of valleys) can be written conveniently in matrix notation as $%
\delta \boldsymbol{v}=\boldsymbol{D\epsilon }$, where $\left\{ \delta 
\boldsymbol{v}\right\} _{l}=\delta v_{l}$ and $\boldsymbol{D}$ is the
deformation potential $L\times 6$ matrix (containing the deformation
potential coupling constants). $\boldsymbol{\epsilon }=\left[ 
\begin{array}{cccccc}
\epsilon _{xx} & \epsilon _{yy} & \epsilon _{zz} & \epsilon _{xy} & \epsilon
_{xz} & \epsilon _{yz}%
\end{array}%
\right] ^{T}$ is the strain component vector and $\epsilon _{\alpha \beta }=%
\frac{1}{2}(\partial u_{\alpha }/\partial \alpha +\partial u_{\beta
}/\partial \beta )$ are the symmetric strain components of displacement $%
\boldsymbol{u}$. For example, for the six Si conduction band minima (see
Fig. \ref{Si_band}) we have $\delta v_{l}=\Xi _{d}(\varepsilon
_{xx}+\varepsilon _{yy}+\varepsilon _{zz})+\Xi _{u}\varepsilon _{ll}$ \cite%
{herring:1956}, where $\Xi _{d}$ ($\Xi _{u}$) is the dilatational (uniaxial)
deformation potential constant. 

Here we deal with long wavelength limit where the phonon field can be
identified with a classical acoustic wave $\boldsymbol{u}=\boldsymbol{e}%
u\exp [-i(\boldsymbol{q\cdot r}-\omega t)]$ with polarization $\boldsymbol{e}
$ ($\left\vert \boldsymbol{e}\right\vert =1$). The strain now reduces to $%
\epsilon _{\alpha \beta }=\frac{-i}{2}(\widehat{q}_{\beta }e_{\alpha }+%
\widehat{q}_{\alpha }e_{\beta })qu$ ($\widehat{\boldsymbol{q}}=\boldsymbol{q/%
}\left\vert \boldsymbol{q}\right\vert $), which can be expressed in matrix
form as $\boldsymbol{\epsilon }=-iq\boldsymbol{Su}$ and we find equation%
\begin{equation}
\delta \boldsymbol{v}=-iq\boldsymbol{DSu},  \label{eq:Si_defor}
\end{equation}
which describes  how the displacement makes the band edges oscillate in a
many valley system.\ In the diffusive long wavelength limit the phonon
momentum itself cannot transfer the electrons from one minima to another,
because this process would require large momentum $q\sim 1/a$ ($a$ is the
lattice constant). Then the electron nonequilibrium, generated by the
acoustic field, relaxes towards local equilibrium by two processes:
diffusion and elastic intervalley impurity scattering. When the strain lifts
the valley degeneracy elastic intervalley scattering provides a path for the
electron system to relax towards local equilibrium. This path is favorable
if the time scale related to diffusion over length $\sim q^{-1}$is
sufficiently large, i.e., when $q^{2}D<1/\tau _{iv},$ where $1/\tau _{iv}$\
is the total elastic intervalley transition rate ($D$ is the diffusion
coefficient). In this limit the linearized many-valley relaxation-time
Boltzmann equation \cite{kragler:1980} reduces to a simple rate equation,
which couples the change in the electron density $\delta n_{l}$ of valley $l$
to that of valley $m$ via intervalley scattering rate $\tau _{lm}^{-1}$: 
\begin{equation}
-\frac{\partial \delta n_{l}}{\partial t}=\sum\nolimits_{m}\tau _{lm}^{-1}%
\left[ \delta n_{l}-\delta n_{m}-2\nu _{1}(\varepsilon _{F})(\delta
v_{l}-\delta v_{m})\right] .  \label{dn_rate}
\end{equation}%
Here $\nu _{1}(\varepsilon _{F})$ is the single spin and valley density of
states at Fermi level  $\varepsilon _{F}\gg k_{B}T$. We assume that strain
equivalent valleys are coupled with rate $\tau _{lm}^{-1}=\tau _{1}^{-1}$
and that the valleys whose degeneracy can be lifted with strain are coupled
with rate $\tau _{lm}^{-1}=\tau _{2}^{-1}$. In the case of Si the coaxial
valleys are always equivalent whereas the degeneracy of the perpendicular
valleys can be lifted (see Fig. \ref{Si_band}). Now the solution of Eq. (\ref%
{dn_rate}) is $\delta \boldsymbol{n}=2\upsilon _{1}(\varepsilon
_{F})(1+i\omega \tau _{iv})^{-1}\boldsymbol{M}\delta \boldsymbol{v}$, where $%
\tau _{iv}=\tau _{2}/L$ and $\left\{ \boldsymbol{M}\right\} _{l,m}=\delta
_{l,m}-L^{-1}$. The phonon-electron relaxation rate $1/\tau _{\boldsymbol{q}}
$ is related to the dissipated heat $Q$ of the acoustic field through
standard relation $1/\tau _{\boldsymbol{q}}=Q/J_{\varepsilon }=\omega
\left\langle \text{Im}\left\{ \delta \boldsymbol{v}\right\} \cdot \text{Re}%
\left\{ \delta \boldsymbol{n}\right\} \right\rangle /J_{\varepsilon }$,
where $J_{\varepsilon }$ is the acoustic energy flux density and $%
\left\langle \text{ \ \ }\right\rangle $ stands for time average. Using this
relation and Eq. (\ref{eq:Si_defor}) we find  
\begin{equation}
\left( \tau _{\boldsymbol{q}\lambda }\right) ^{-1}=\frac{2\nu
_{1}(\varepsilon _{F})}{\rho _{d}v_{\lambda }^{2}}\frac{\omega _{\boldsymbol{%
q}\lambda }^{2}\tau _{iv}}{1+\omega _{\boldsymbol{q}\lambda }^{2}\tau
_{iv}^{2}}\Phi _{\lambda },  \label{eq:tau_phe}
\end{equation}%
where we have used linear dispersion relations $\omega _{\boldsymbol{q}%
\lambda }=v_{\lambda }q$ ($\lambda $\ is the mode index). The factor $\Phi
_{\lambda }=\boldsymbol{e}^{T}\boldsymbol{S}^{T}\boldsymbol{D}^{T}%
\boldsymbol{MDS}\boldsymbol{e}$ and it obviously depends only on the
polarization $\boldsymbol{e}$, on the direction of propagation $\widehat{%
\boldsymbol{q}}$ and on the deformation potential coupling constants. In the
case of Si we have $\Phi _{\lambda }=2\Xi _{u}^{2}\left[ \sum_{i}(\widehat{q}%
_{i}e_{i})^{2}-\frac{1}{3}(\widehat{\boldsymbol{q}}\cdot \boldsymbol{e})^{2}%
\right] $ and $1/\tau _{iv}=$ $6/\tau _{2}$. Note that Eq. (\ref{eq:tau_phe}%
) does not depend on screening, because there are no total electron density
fluctuations, i.e., \  $\sum \delta n_{l}=0$.

We can describe a degenerate electron system by an equilibrium distribution
at temperature $T_{e}$. This holds even in the presence of net heat flow
between electrons and phonons. The heat flow only creates a non-equilibrium
between the electrons and phonons, which relaxes towards equilibrium at rate 
$1/\tau _{\boldsymbol{q}\lambda }$ per single phonon mode. By following
Perrin and Budd \cite{perrin:1972} this non-equilibrium can be expressed
using the relaxation time approximation of the phonon-electron collision
integral
\begin{equation}
\left( \frac{\partial N(\omega _{\boldsymbol{q}\lambda })}{\partial t}%
\right) _{ph-e}=-\frac{N(\omega _{\boldsymbol{q}\lambda })-N_{T_{e}}(\omega
_{\boldsymbol{q}\lambda })}{\tau _{\boldsymbol{q}\lambda }},  \label{eq:dNdt}
\end{equation}%
where $N(\omega _{\boldsymbol{q}\lambda })$ and $N_{T}(\omega _{\boldsymbol{q%
}\lambda })=\left[ 1-\exp (\hbar \omega _{\boldsymbol{q}\lambda }/k_{B}T)%
\right] ^{-1}$ are the nonequilibrium and equilibrium phonon distribution
functions, respectively. The total stationary heat flow $P$ through the
coupled electron-phonon system is the energy average of the collision
integral:
\begin{equation}
P=\sum_{\lambda }\int \frac{d\boldsymbol{q}}{(2\pi )^{3}}\hbar \omega _{%
\boldsymbol{q}\lambda }\left( \frac{\partial N(\omega _{\boldsymbol{q}%
\lambda })}{\partial t}\right) _{ph-e}\text{,}  \label{eq:Pdef}
\end{equation}%
where the summation is performed over the acoustic eigenmodes of the
crystal. The only experimentally meaningful situation is such that the
phonon system is coupled to some thermalizing bath, which is at temperature $%
T_{0}$. If the coupling is strong or $P$ \ is small we can approximate $%
N(\omega _{\boldsymbol{q}\lambda })\simeq N_{T_{ph}}(\omega _{\boldsymbol{q}%
\lambda })$, where $T_{ph}$ is the (possibly local) phonon temperature, and
Eq. (\ref{eq:Pdef}) reduces to the familiar form:
\begin{equation}
P=F(T_{e})-F(T_{ph}),  \label{eq:PF}
\end{equation}%
where $F(T)$ is the energy flow rate control function. \ Using Eqs. (\ref%
{eq:tau_phe})-(\ref{eq:PF}) and assuming that $(k_{B}T/\hbar \tau
_{iv}^{-1})^{2}$ is clearly below unity the energy flow rate control
function can be expressed in a closed form%
\begin{align}
F(T)& =\frac{\nu _{1}(\varepsilon _{F})B_{5}}{\pi ^{2}\rho _{d}\hbar ^{5}}%
\sum_{\lambda }\left\langle \frac{\Phi _{\lambda }}{v_{\lambda }^{5}}%
\right\rangle _{\Omega }\tau _{iv}\left( k_{B}T\right) ^{6}  \notag \\
& =\frac{2\nu _{1}(\varepsilon _{F})\Xi _{u}^{2}B_{5}}{45\pi ^{2}\rho
_{d}\hbar ^{5}v_{T}^{5}}\left[ \frac{17}{8}+\left( \frac{v_{T}}{v_{L}}%
\right) ^{5}\right] \tau _{2}\left( k_{B}T\right) ^{6},  \label{eq:F(T)_Si}
\end{align}%
where the first equality is valid for arbitrary many-valley system. The
constant $B_{5}=\int_{0}^{\infty }dxx^{5}/\left[ 1-\exp (x)\right] =120\pi
^{6}/945$ and $\left\langle \text{ \ }\right\rangle _{\Omega }$ stands for
average over a solid angle. The second equality applies for silicon and
there we have further assumed that the phonon eigenmodes are isotropic and
that they are described by the longitudinal and transversal sound velocities 
$v_{L}$ and $v_{T}.$

Eq. (\ref{eq:F(T)_Si}) is valid when $(k_{B}T/\hbar \tau _{iv}^{-1})^{2}<1$
and $q_{T}^{2}D<1/\tau _{iv}$. At low temperatures the dominating condition
is the latter and can be written also as $q_{T}l\sqrt{\tau _{iv}/\tau }<1$,
where $\tau $ is the momentum relaxation time. Condition $q_{T}l\sqrt{\tau
_{iv}/\tau }=1$ defines the crossover temperature below which elastic
intervalley scattering induced electron-phonon relaxation dominates over
diffusion. If $\tau _{iv}$ is not orders of magnitude larger than $\tau $
this differs very little from the impure-pure threshold $q_{T}l=1$.

Eq. (\ref{eq:F(T)_Si}) suggests that intervalley scattering induced
electron-phonon energy relaxation rate $\tau _{\epsilon }^{-1}\propto \tau
_{iv}T_{e}^{4}$, which can be seen from approximate rate equation $%
dP/dT_{e}\approx C_{e}\tau _{\epsilon }^{-1}$, where $C_{e}=\gamma T_{e}$ is
the electron heat capacity. As the phonon mediated dephasing rate $1/\tau
_{i}^{ph}\propto \tau _{\epsilon }^{-1}$\cite{bergmann:1990} we find an
important relation $1/\tau _{i}^{ph}\propto \tau _{iv}T_{e}^{4}$.

As already pointed above screening plays no role in $1/\tau _{\boldsymbol{q}%
\lambda }$ and as a result intervalley scattering induced electron-phonon
energy flow rate in Eq. (\ref{eq:F(T)_Si}) does not include any screening
parameters, like for example screening wave vector $\kappa .$ Note, however,
that there exists also single-valley contribution to the energy relaxation
which is due to number density fluctuations in a single valley, but this
conribution is strongly screened in doped semiconductors \cite{sota:1986}.
By using the single valley result calculated by Sergeev \textit{et al.} \cite%
{sergeev:2005} and Eq. (\ref{eq:F(T)_Si}) we find that the ratio between
many-valley and single valley energy flow rate scales roughly as $\sim
1500\left( l\kappa \right) ^{2}\left( \tau _{iv}/\tau \right) (\kappa /T)^{2}
$, where $[\kappa ]=$ nm$^{-1}$and $[T]$=K. Thus the many-valley effect is
expected to fully dominate in the diffusive limit at high electron densities
and low temperatures. We have tested Eq. (\ref{eq:F(T)_Si}) experimentally
in the case of n$^{+}$ Si:

\begin{table}[tbp]
\caption{The characteristics of the samples: $N$ - carrier concentration, $%
\protect\rho _{e}$ - 1.5 K resistivity, $l$ - electron mean free path, $d$ -
n$^{+}$ Si film thickness. All samples have 400 nm thick buried oxide layer.}
\label{table1}%\label{table1}
%\footnotetext[1]{VTT.}
\begin{ruledtabular}
\begin{tabular}{ccccc}
sample & $N$ ($10^{25}$m$^{-3}$) & $\rho _{e}$ ($10^{-5}\Omega \cdot $ m) & $%
l$ (nm) & $d$ (nm) \\ \hline
A & 3.5 & 1.04 & 5.06 & 70 \\ 
F & 6.7 & 0.63 & 5.42 & 58 \\ 
G & 12.0 & 0.51 & 4.54 & 58 \\ 
H & 16.0 & 0.44 & 4.34 & 58%
\end{tabular}
\end{ruledtabular}
\end{table}

\begin{figure}[b]
\begin{center}
\includegraphics[width=75mm,height=!]{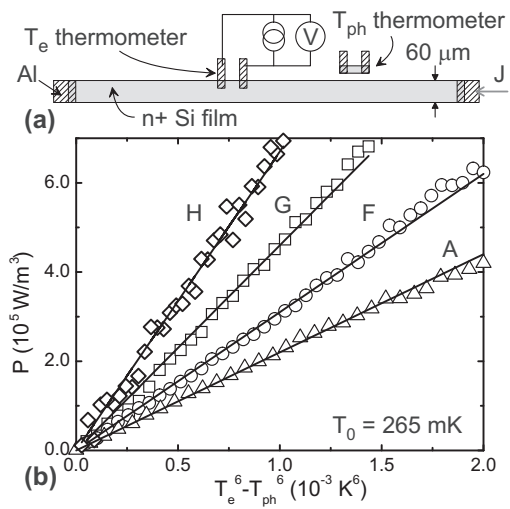}
\end{center}
\par
%\label{scem_T6}
\caption{(a) Schematic illustration of the sample geometry and the
measurement setup. The $\sim $9500 $\protect\mu $m long n$^{+}$ Si film is
heated with a DC current density $J$. $T_{e}$ and $T_{ph}$ are measured
using current biased S-Sm-S (Al-Si-Al) contacts (only the biasing circuit
for $T_{e}$ is depicted). $T_{ph}$ thermometer is electrically isolated from
the main Si film by a $\sim $1 $\protect\mu $m gap. (b) The power density $P=%
\protect\rho _{e}J^{2}$ vs. $T_{e}^{6}-T_{ph}^{6}$ for samples with
different carrier concentrations at bath temperature of 265 mK. }
\label{scem_T6}
\end{figure}

The n$^{+}$ Si samples were fabricated on unibond silicon-on-insulator
substrates. Properties of the samples are listed in Table \ref{table1} and a
detailed description about the sample fabrication can be found in \cite%
{prunnila:2002}. The sample geometry and the experiment is depicted in Fig. %
\ref{scem_T6}(a). In the experiments the samples were mounted on a sample
holder of a dilution refrigerator. The electron and phonon temperatures were
simultaneously measured by utilizing the
superconductor-semiconductor-superconductor (S-Sm-S) thermometry \cite%
{savin:2001} while the electron gas in the Si film was heated with a DC
power density $P=\rho _{e}J^{2}$ created by electric current density $J$.
Note that as the electronic coupling to the n$^{+}$ Si film is made via
superconducting Al the heat flow in the experiment follows accurately a path
electrons$\rightarrow $phonons$\rightarrow $substrate/sample holder
(phonons) and, therefore, the experimental $P$ is equal to the
left-hand-side of Eq. (\ref{eq:Pdef}). Heating of the electron gas can cause
a substantial increase in the temperature of the phonon thermometer, as
reported recently for a similar n+ Si sample as discussed here \cite%
{kivinen:2003}. To assure that the nonequilibrium phonon distribution (of
the phonons that interact with the electrons in the Si layer) can be
reasonably described with an equilibrium distribution function we consider
heating power range where $(T_{ph}-T_{0})/T_{0}$ is clearly below unity.

\begin{figure}[t]
\begin{center}
\includegraphics[width=75mm,height=!]{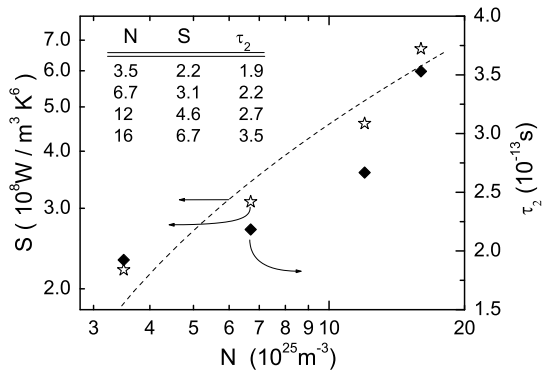}
\end{center}
\par
%\label{FperT6}
\caption{Slopes $S$ of the linear fits in Fig. \protect\ref{scem_T6}(b) and
intervalley scattering time $\protect\tau _{2}$ [determined from $S$ and Eq.
(\protect\ref{eq:F(T)_Si})] as a function of electron density $N$. The
dashed curve is a polynomial fit that serves as a guide for the eye. The
inset shows tabulated values of $S$ and $\protect\tau _{2}$ (in the units of
the axes). }
\label{FperT6}
\end{figure}

Fig. \ref{scem_T6}(b) shows the experimental power density vs. $%
T_{e}^{6}-T_{ph}^{6}$ at bath temperature $T_{0}\approx 265$ mK. The solid
curves are least of square fits to $P=S(T_{e}^{6}-T_{ph}^{6})$ with the
slope $S$ as a single fitting parameter. We observe that the electron-phonon
temperature response predicted by Eq. (\ref{eq:F(T)_Si}) describes all the
samples extremely well. The slopes $S$ are plotted against the electron
density in Fig. \ref{FperT6} (left vertical axis). $S$ increases as function
of $N$, which is expected result, because $F(T)/T^{6}\propto $ $\nu
_{1}(\varepsilon _{F})$.

In order to perform more quantitative comparison between the theory and
experiment we estimate the density of states from free electron gas
expression $\nu _{1}(\varepsilon _{F})=\nu _{1}^{0}(\varepsilon _{F})=\left(
m_{de}/2\pi \hbar ^{2}\right) \left( 3\pi ^{2}N/L\right) ^{1/3}$, where we
use Si density of states mass $\ m_{de}=0.322m_{e}$ ($m_{e}$ is the free
electron mass). For the other parameters we use the typical values for Si: $%
\Xi _{u}=9.0$ eV, $\rho _{d}=2.33\times 10^{3}$ kgm$^{-3},v_{T(L)}=4700$ $%
(9200)$ m/s.\ Now the intervalley scattering time $\tau _{2}=6\tau _{iv}$
can be determined from $S=F(T)/T^{6}$ [see Eq. (\ref{eq:F(T)_Si})] and it is
plotted on the right vertical axis of\ Fig. \ref{FperT6}. The cross-over
temperature from the condition $q_{T}l\sqrt{\tau _{iv}/\tau }=1$ is found to
be $\approx $ 5 K (average from all the samples). Thus we are at $q_{T}l%
\sqrt{\tau _{iv}/\tau }\ll 1$ limit.

Eq. (\ref{eq:tau_phe}) gives also the phonon or ultrasonic attenuation
constant $\alpha _{\boldsymbol{q}\lambda }=$ $\left( \tau _{\boldsymbol{q}%
\lambda }\right) ^{-1}/(2v_{\lambda })$. Using this result and the
ultrasonic attenuation data obtained by M. Dutoit \cite{dutoit:1970} from n$%
^{+}$ Si with $N=2\times 10^{25}$ m$^{-3}$ at temperature of 2 K $\ $at $%
\omega =1.48\times 10^{9}$ 1/s we find $\tau _{2}(N=2\times 10^{25}$ m$%
^{-3})\approx 0.3$ ps. This fits to our measurements extremely well, which
is an important result: experiment that probes heat transport between
electrons and one coherent acoustic mode \cite{dutoit:1970} coincides with
our experiment that probes heat transport between electrons and phonon gas
obeying quantum statistics.

At high $N$ one would expect slowly decreasing or a roughly constant $\tau
_{2},$ while our results show a weak increase as a function of $N$. This
unexpected result could be explained by noting that our samples are in the
limit of strong disorder ($k_{F}l\approx 3.6$ on average from Table \ref%
{table1}). Whereas, Eq. (\ref{eq:F(T)_Si}) is essentially based on a
semiclassical free electron gas model, at least finally when the
approximation $\nu _{1}(\varepsilon _{F})=\nu _{1}^{0}(\varepsilon _{F})$ is
made. Correction terms arising from interaction and quantum interference
effects can be included to our model in the spirit of Ref. \cite{sota:1986}.
As similar terms appear in the conductivity the magnitude of these quantum
corrections can be estimated from low field magnetoresistance and
temperature dependency of resistivity. At the moment such data for thin film
n$^{+}$ Si is not available.

Finally, we point out that the intervalley scattering induced
electron-phonon energy relaxation can be observed also in several other
material systems than n$^{+}$ Si. Canonical examples would be n$^{+}$ Ge and
two dimensional electron gas in $\left( 111\right) $ Si inversion layer. As
the $\Gamma $-point in the valence band of elemental semiconductors is
divided into heavy hole, light hole and split-off bands the effect should be
particularly strong in various hole systems. However, due to complicated
nature of the valence band maximum and effectively zero distance of the
different bands in k-space the theory, which is valid for conduction band
electrons, should be modified\cite{sota:1984}.

In summary, we have studied the effect of elastic intervalley transitions on
the electron-phonon energy relaxation rate in many-valley semiconductors in
the diffusive limit. We derived a general expression for the electron-phonon
energy flow rate [Eq. (\ref{eq:F(T)_Si})] and discussed the special case of n%
$^{+}$ silicon. Low temperature experiments on heavily doped Si samples were
performed and good agreement between the theory and the experiment was found.

\begin{acknowledgments}
We want to acknowledge the skillful contribution of M. Markkanen in the
sample fabrication. This work has been partially funded by the Academy of
Finland (project numbers 46804, 205470, 205467 and 53903). PK also
acknowledges financial support of Ulla Tuominen and Emil Aaltonen
Foundations.
\end{acknowledgments}

%TCIMACRO{\QSubDoc{Include dummy_1}{\input{dummy_1.tex}}}%
%BeginExpansion
%\input{dummy_1.tex}%
%EndExpansion

%\input{dummy.tex}

%\newpage \textbf{Figure captions}

%\begin{thebibliography}{99} %  
%\bibliographystyle{ieeetr}

\bibliographystyle{apsrev}
\bibliography{EP_refs_mpr_2}

%\newpage %\begin{figure}[h]

%\begin{center}
%\includegraphics[width=84mm,height=!]{Fig1}

%\textbf{Fig.1. M. Prunnila, Applied Physics Letters}
%\end{center}

\end{document}